\documentclass[showpacs,prd,aps,10pt,superscriptaddress,altaffilletter,floatfix,twocolumn]{revtex4-1}
\usepackage{graphicx}
\usepackage{amsmath}
\usepackage{amssymb}
\usepackage{color}
\usepackage{xspace}
\usepackage{dcolumn}
\usepackage{soul}
\usepackage{multirow}
\usepackage[colorlinks, linkcolor=blue, anchorcolor=blue, urlcolor = blue, citecolor=blue]{hyperref}
\newcolumntype{C}[1]{>{\centering\let\newline\\\arraybackslash\hspace{0pt}}m{#1}}

\usepackage{url}

\RequirePackage{lineno}

\begin{document}

\newcommand{\isot}[2]{$^{#2}$#1}
\newcommand{\isotbold}[2]{$^{\boldsymbol{#2}}$#1}
\newcommand{\xeiso}{\isot{Xe}{136}\xspace}
\newcommand{\thsrc}{\isot{Th}{228}\xspace}
\newcommand{\cosrc}{\isot{Co}{60}\xspace}
\newcommand{\rasrc}{\isot{Ra}{226}\xspace}
\newcommand{\cssrc}{\isot{Cs}{137}\xspace}
\newcommand{\betascale}  {$\beta$-scale}
\newcommand{\kevkgyr}  {keV$^{-1}$ kg$^{-1}$ yr$^{-1}$}
\newcommand{\nonubb}  {$0\nu \beta\!\beta$\xspace}
\newcommand{\nonubbbf}  {$\boldsymbol{0\nu \beta\!\beta}$\xspace}
\newcommand{\twonubb} {$2\nu \beta\!\beta$\xspace}
\newcommand{\es}{$0^+_1$\xspace}
\newcommand{\bb} {$\beta\!\beta$\xspace}
\newcommand{\vadc} {ADC$_\text{V}$}
\newcommand{\uadc} {ADC$_\text{U}$}
\newcommand{\mus} {\textmu{}s}
\newcommand{\chisq} {$\chi^2$}
\newcommand{\mum} {\textmu{}m}
\newcommand{\red}[1]{{\xspace\color{red}#1}}
\newcommand{\blue}[1]{{\xspace\color{blue}#1}}
\newcommand{\RunTwoA}{Run 2a}
\newcommand{\RunTwo}{Run 2}
\newcommand{\RunTwoBC}{Runs 2b and 2c}
\newcommand{\SP}[1]{\textsuperscript{#1}}
\newcommand{\SB}[1]{\textsubscript{#1}}
\newcommand{\SPSB}[2]{\rlap{\textsuperscript{#1}}\SB{#2}}
\newcommand{\pmasy}[3]{#1\SPSB{$+$#2}{$-$#3}}
\newcommand{\matel}{$M^{2\nu}$}
\newcommand{\psfac}{$G^{2\nu}$}
\newcommand{\tbeta}{T$_{1/2}^{0\nu\beta\beta}$}
\newcommand{\exolimit}[1][true]{\pmasy{2.6}{1.8}{2.1}$ \cdot 10^{25}$}
\newcommand{\exomeasurement}{\tbeta{}= \exolimit{}~yr}
\newcommand{\U}{\text{U}}
\newcommand{\V}{\text{V}}
\newcommand{\X}{\text{X}}
\newcommand{\Y}{\text{Y}}
\newcommand{\Z}{\text{Z}}
\newcommand{\bqcm}{${\rm Bq~m}^{-3}$}
\newcommand{\nonunorm}{N_{{\rm Err, } 0\nu\beta\beta}}
\newcommand{\nonunum}{n_{0\nu\beta\beta}}
\newcommand{\cussim}[1]{$\sim$#1}
\newcommand{\halflife}[1]{$#1\cdot10^{25}$~yr}
\newcommand{\numspec}[3]{$N_{^{#2}\mathrm{#1}}=#3$}
\newcommand{\TD}[1]{\textcolor{red}{#1}}
\newcommand{\TB}[1]{\textcolor{blue}{#1}}
\newcommand{\FIXME}[1]{\textcolor{red}{#1}}
\newcommand{\PI}{Phase~I\xspace}
\newcommand{\PII}{Phase~II\xspace}
\newcommand{\Rn}{radon\xspace}
\newcommand\Tstrut{\rule{0pt}{2.6ex}} 

\setstcolor{blue}

\title{Search for Two-neutrino Double-Beta Decay of $^{136}$Xe to the \es excited state of $^{136}$Ba with the Complete EXO-200 Dataset}

\author{S.~Al~Kharusi}\affiliation{Physics Department, McGill University, Montreal, Quebec H3A 2T8, Canada}
\author{G.~Anton}\affiliation{Erlangen Centre for Astroparticle Physics (ECAP), Friedrich-Alexander-University Erlangen-N\"urnberg, Erlangen 91058, Germany}
\author{I.~Badhrees}\altaffiliation{Permanent address: King Abdulaziz City for Science and Technology, Riyadh, Saudi Arabia}\affiliation{Physics Department, Carleton University, Ottawa, Ontario K1S 5B6, Canada}
\author{P.S.~Barbeau}\affiliation{Department of Physics, Duke University, and Triangle Universities Nuclear Laboratory (TUNL), Durham, North Carolina 27708, USA}
\author{D.~Beck}\affiliation{Physics Department, University of Illinois, Urbana-Champaign, Illinois 61801, USA}
\author{V.~Belov}\affiliation{Institute for Theoretical and Experimental Physics named by A.I. Alikhanov of National Research Centre ``Kurchatov Institute'', Moscow 117218, Russia}\altaffiliation[Now a division of National Research Center ``Kurchatov Institute'', Moscow 123182, Russia]{}
\author{T.~Bhatta}\altaffiliation{Present address: Department of Physics and Astronomy, University of Kentucky, Lexington, Kentucky 40506, USA}\affiliation{Department of Physics, University of South Dakota, Vermillion, South Dakota 57069, USA}
\author{M.~Breidenbach}\affiliation{SLAC National Accelerator Laboratory, Menlo Park, California 94025, USA}
\author{T.~Brunner}\affiliation{Physics Department, McGill University, Montreal, Quebec H3A 2T8, Canada}\affiliation{TRIUMF, Vancouver, British Columbia V6T 2A3, Canada}
\author{G.F.~Cao}\affiliation{Institute of High Energy Physics, Beijing 100049, China}
% ...
\collaboration{EXO-200 Collaboration}

\date{\today}

\begin{abstract}
A new search for two-neutrino double-beta (\twonubb) decay of 
$^{136}$Xe to the \es excited state of $^{136}$Ba is performed 
with the full EXO-200 dataset. A deep learning-based convolutional neural network is used to discriminate signal from background events. Signal detection efficiency is increased relative to previous searches by EXO-200 by more than a factor of two. With the addition of the \PII dataset taken with an upgraded detector, the median 90\% confidence level half-life sensitivity of \twonubb decay to the \es state of $^{136}$Ba is $2.9 \times 10^{24}$~yr using a total $^{136}$Xe exposure of 234.1~kg~yr. No statistically significant evidence for \twonubb decay to the \es state is observed, leading to a lower limit of $T^{2\nu}_{1/2}(0^+ \rightarrow 0^+_1) > 1.4\times10^{24}$~yr at 90\% confidence level, improved by 70$\%$ relative to the current world's best constraint.
\end{abstract}


\maketitle

\section{Introduction}
Double-beta ($\beta\beta$) decay is a second-order weak transition in which two neutrons simultaneously decay into two protons. The observation of the decay mode without neutrino emission (\nonubb) would demonstrate the Majorana nature of neutrinos, which is a fundamental question in particle physics. The current best half-life limits on \nonubb set by various experiments \cite{exo_prl_2019, CUORE:2021mvw, GERDA:2020_final_results, KamLAND-Zen:2022tow, Majorana:2022udl} for different isotopes are in the range of 10$^{25-26}$ yr. The $\beta\beta$ mode with two accompanying neutrinos (\twonubb) is a standard model process and has been observed in over ten isotopes \cite{Barabash:2020nck_2vbb_review} with half-life between 10$^{19}$ and 10$^{24}$ yr. The \twonubb can also decay to the excited states of the daughter nucleus. These $\beta\beta$ modes share the same initial and final nucleus as the \nonubb decay, and might have correlated nuclear matrix elements (NMEs) \cite{PhysRevC.106.054302_NMEs1, Jokiniemi:2022ayc_NMEs2}. Measuring the decays to the excited states offers additional experimental input to the calculation of \twonubb NMEs, which might help reduce the theoretical uncertainties for \nonubb NMEs. 

The \twonubb decay to the excited states of the daughter nucleus are suppressed by orders of magnitude with respect to the decay to the ground state because of the reduced phase space from smaller Q values \cite{exo_es_2016}. In addition, the decays to the $2^+$ states are highly suppressed by angular momentum. Therefore, the decay to the first excited state of the $0^+$ state, denoted as $0^+_1$ hereafter, is the most viable. The first experimental limit derived for the decay to the excited state dates back to 1977 for $^{76}$Ge \cite{1977_ge76}. The first positive signal was observed for $^{100}$Mo decay to the $0^+_1$ excited state of $^{100}$Ru in 1995 \cite{Barabash:1995fn_Mo100}, followed by confirmations from a series of other works \cite{Kidd:2009_Mo100, CUPID-Mo:2022cel}. The $^{150}$Nd decay to the $0^+_1$ excited state of $^{150}$Sm is the only other isotope where a positive signal has been observed \cite{Barabash:2004_Nd150}. Stringent limits have been set for other isotopes \cite{barabash2017double_es_sum}. The CUORE experiment placed a limit of $T^{2\nu}_{1/2}(0^+ \rightarrow 0^+_1) > 2.4 \times 10^{23}$~yr for $^{130}$Te in 2019 \cite{CUORE:2018ncg_ES}. More recently, the MAJORANA-DEMONSTRATOR placed a limit of $T^{2\nu}_{1/2}(0^+ \rightarrow 0^+_1) > 7.5 \times 10^{23}$~yr for $^{76}$Ge \cite{MAJORANA:2020shy_mjd_es}. KamLAND-Zen and EXO-200 performed measurements for $^{136}$Xe~\cite{KamLAND-Zen:2015tnh_kamland_es, exo_es_2016}, and the current best limit is set at $T^{2\nu}_{1/2}(0^+ \rightarrow 0^+_1) > 8.3\times10^{23}$~yr. 

The $^{136}$Xe $\beta\beta$ decay to the ground state of $^{136}$Ba has a Q value of 2457.83 keV \cite{Mccutchan:2018jsr_A136}. The $\beta\beta$ decay to the \es state of $^{136}$Ba has a Q value of 878.8 keV, followed by the emission of two characteristic de-excitation $\gamma$'s with energies of 760.5 keV and 818.5 keV from consecutive de-excitations from the $0^+_1 \rightarrow 2^+_1 \rightarrow 0^+$ states of $^{136}$Ba \cite{Mccutchan:2018jsr_A136}. Two electrons accompanied by two $\gamma$'s provide distinctive event signatures. The theoretically predicted half-life for \twonubb decay of $^{136}$Xe to \es excited state of $^{136}$Ba ranges from 10$^{23}$ to 10$^{26}$ yr using different nuclear models \cite{Jokiniemi:2022_es_theory_latest,theory_predict}. The sensitivities of current generation experiments are approaching this level, presenting a chance for discovery. In this paper, we report a new search for the decay to \es state of $^{136}$Ba using the complete EXO-200 dataset.

\section{The EXO-200 experiment}
The EXO-200 detector is a single phase liquid xenon (LXe) time projection chamber (TPC). The experiment was located in the Waste Isolation Pilot Plant (WIPP) near Carlsbad, New Mexico~(USA), with an overburden of 1624$^{+22}_{-21}$ meters of water equivalent~\cite{EXO_cosmogenics}. EXO-200 operated between Sept. 2011 and Dec. 2018 in two phases. \PI data taking was halted in Feb. 2014 due to underground incidents at the host facility unrelated to the experiment. \PII operation began in May 2016 after underground access was restored. Among other things, the detector electronics were upgraded before \PII to reduce coherent noise on the avalanche photodiodes (APDs) during the forced outage. A brief review of the detector is given in this section. More details on the experiment can be found elsewhere \cite{Auger:2012gs,Albert2013,EXO-200:2021srn_part2}.

\subsection{Detector description}
The EXO-200 TPC was housed in a cylindrical, thin-walled copper vessel with a cathode in the middle, resulting in two back-to-back drift regions with $\sim$18~cm radius and $\sim$20~cm drift length. Two wire planes crossing of 60$^{\circ}$ are placed at each anode end of the drift. The induction plane (V-wires) records the transit of charge, while the collection plane (U-wires) collects the charge. The scintillation photons produced simultaneously with the ionization were detected by arrays of large area avalanche photodiodes (LAAPDs)~\cite{Neilson:2009kf} behind the anode planes. The entire copper vessel enclosing the TPC was submerged in HFE-7000 cryofluid~\cite{m3m}, enclosed by additional layers of passive shield made of 5.4~cm copper and 25~cm lead in all directions~\cite{Auger:2012gs}. An active muon veto was composed of scintillator panels, covering four sides of the detector, providing $>$94\% tagging efficiency for muons passing through the TPC~\cite{EXO_cosmogenics}.

\subsection{Data and simulation}
Event reconstruction makes use of both the light and charge signals produced by particle interactions in the LXe. Signals from U-wires and V-wires are used to extract the position of the energy deposit in the plane normal to the electric field direction. The time difference between the registered light signal on the LAAPDs and the charge collection on U-wires is used to obtain the $z$ position along the drift field, using the measured electron drift velocity~\cite{EXO_diffusion}. This allows full three dimensional (3D) vertex reconstruction for individual energy deposits, defined as charge clusters in the analysis to distinguish them from the true energy deposits. Small charge deposits with energy below the V-wire but above the U-wire detection threshold do not have a reconstructed $xy$ position. Events reconstructed with multiple charge clusters are referred to as ``multi-site'' (MS) while single charge cluster events are denoted as ``single-site'' (SS). The SS/MS identification can help distinguish $\beta\beta$ events from $\gamma$'s, as $\beta$-like events are primarily SS, while $\gamma$'s tend to be MS. Only 12\% (14\%) of $\gamma$ events are reconstructed as SS events near the Q$_{\beta\beta}$ of $^{136}$Xe for the \isot{Th}{228} (\isot{Ra}{226}) source~\cite{exo_prl_2019}.

The event energy is reconstructed by a linear combination of the light energy measured by the LAAPDs and the charge energy measured by the U-wires to fully exploit the anti-correlation between these two channels~\cite{Conti2003}. Such a linear combination cancels out anti-correlated fluctuations in light and charge signals, thus optimizes the energy resolution, which is especially important for the \nonubb search.

A comprehensive Monte Carlo (MC) detector simulation based on GEANT4~\cite{GEANT42006} was developed to model the responses to various signal and background interactions. This simulation models the entire chain from  energy deposits produced by interactions in the LXe to waveforms created on the crossed-wire planes by the ionization propagating through the detector. The simulated waveforms are treated the same way in the reconstruction and analysis framework as the real data. The simulation is benchmarked by external $\gamma$ calibration sources located $\sim$10 cm away from the fiducial volume (FV) at set positions near the cathode and the anodes~\cite{Albert2013}.

\section{Analysis Overview}
\label{sec:analysis overview}
The dataset used in this analysis is the same as that used in \cite{exo_prl_2019}, with a total livetime of 1183.1 days after run quality selections. 
All clusters of each event are required to lie within a FV, defined as a hexagon in the $xy$ plane with an apothem of 162~mm and more than 10~mm away from the cylindrical polytetrafluoroethylene reflector, as well as the cathode and the V-wire planes. This FV contains $3.31\times10^{26}$ atoms of $^{136}$Xe, with an equivalent mass of 74.7~kg. The systematic error of this value is included in the common normalization error, which will be discussed in Sec.~\ref{sec:syst}.
After additional muon veto cuts to reduce cosmogenic events, the total $^{136}$Xe exposure is 117.4 (116.7) kg$\cdot$yr in \PI (\PII).

Compared to the analysis based on the Phase-I dataset in \cite{exo_es_2016}, a major improvement to the signal efficiency is achieved by relaxing the 3D cluster reconstruction requirement. Because the signal detection energy threshold on the V-wires ($\sim$200 keV) is higher than that on the U-wires ($\sim$90 keV), a large number of charge clusters with small energy deposits have no V-wire signals, resulting in incomplete $xy$ positions for these clusters. In the previous search, all clusters were required to have fully reconstructed 3D positions. For this search, we only require the sum energy of all fully reconstructed clusters to be above 60\% of the total event energy~\cite{exo_prl_2019}. The efficiency for signal events increased by a factor of $\sim$2, while the total background events only increased by a factor of $\sim$1.4. In addition, the newly developed background discriminator, utilizing event topologies further improves the signal to background ratio. With a relaxed 3D vertex reconstruction cut,  the signal efficiency increased from 24.5$\pm$3.1\% (25.4$\pm$3.1\%) to 58.0$\pm$2.9\% (58.7$\pm$2.9\%) for Phase I (Phase II). The remaining signal inefficiency is from the fiducial volume and energy cuts. Lowering the energy threshold has only a marginal gain in signal efficiency, which is out weighed by the large increase in background efficiency at those energies. As a result, there is little improvement to the sensitivity from lowering the energy threshold and the same 1000 keV threshold from the previous analysis is used.

A deep learning (DL) method based on a convolutional neural network (CNN)~\cite{lecun1989_cnn} is used to separate excited state signals from backgrounds in this work. Unlike the boosted decision tree (BDT) approach used in \cite{exo_es_2016} where cluster information is combined into high level variables, the individual cluster information is directly provided as input to the CNN. This input contains more complete information about the event, which further improves background discrimination. The details will be discussed in Sec.~\ref{sec:ml}. 

The accurate determination of the energies for clusters is essential to identify the characteristic de-excitation $\gamma$'s. The cluster energy is reconstructed from charge signals only, because the light signals are unresolvable among clusters in an event as they appears as one flash of light in the detector. Previous EXO-200 work already showed charge yield in liquid xenon is energy dependent in the MeV range~\cite{EXO_LXeYield}.
 A calibration curve is derived using SS events from \isot{Cs}{137}, \isot{Co}{60}, \isot{Ra}{226} and \isot{Th}{228} calibration sources. Using all available calibration data spanning the experiment lifetime, the calibrated cluster energy scale uncertainty is determined to be within 1\% above 662~keV. 
 
The fitting framework closely follows previous analyses \cite{EXO_PRL17, exo_prl_2019}. The signal search was performed with a negative log-likelihood (NLL) function to fit simultaneously the SS and MS events to two dimensional (2D) Probability Density Functions (PDFs) of event energy and a DL discriminator for different background and signal components. The relative fraction of SS and MS events for each component is set by MC predicted values and constrained by errors determined from data-MC difference in calibration source data (Sec.~\ref{sec:syst}). PDFs are constructed based on simulation validated by extensive calibration data. Residual discrepancies between data and simulation are taken into account as systematic uncertainties. Gaussian terms are added to the NLL to take into account various systematic errors, which will be discussed in Sec.~\ref{sec:syst}. 

A profile likelihood scan was performed to derive the 90\% confidence level (CL) limit at the negative logarithm of the likelihood ratio between a given number of counts and the best fit ($\Delta$NLL) value of 1.35, under the assumption of Wilks’ theorem \cite{wilks1938, cowan1998statistical} considering the large statistics of the dataset in the region of interest. A sensitivity study is performed to compare different analyses. The sensitivity is evaluated by generating a set of toy datasets based on the background model derived from a fit to the energy spectrum of the low background data without the excited state signal included in the PDFs. It represents our best $a~priori$ understanding of the background model.  The 90$\%$ CL limit is derived for each individual toy dataset, and the median of the limit distribution from all toy datasets is defined as the sensitivity. Though the full EXO-200 dataset has been unblinded previously, we avoided tuning the analysis against the low background dataset used to search for excited state signals. The analysis strategy is chosen prior to the final fit based on these sensitivity studies which only rely on calibration data and simulations.

\section{Background discrimination with deep learning}
\label{sec:ml}
The excited state signals have distinct features due to the two accompanying $\gamma'$s of specific energies. The individual cluster energy and position variables for each event are used as inputs for background discrimination, as they are expected to maintain more information about the energy and topology of the $\beta$ and $\gamma$ related interactions. A DL based method was utilized to extract the correlation between both energy and position maintained by the two de-excitation $\gamma'$s from this more complicated input. The details are given in this section.

\subsection{Training dataset and input variables}
\label{sec:input_variables}
To train the background discriminator, the background model obtained from an energy-only fit to the low background data is used to represent the background compositions in the data. The relative fractions of different components of the background model are given by the best fit values. In total, 1.8 million MC events are used to train the network, composed of half signal events and half background events. During the training process, 80\% of the sample was used for training and the remaining 20\% for validation, Training and validation of the background discriminator is done in python using the Pytorch package~\cite{paszke2019pytorch}. 

\begin{figure}[bp!]
    \centering
    \includegraphics[width=\linewidth]{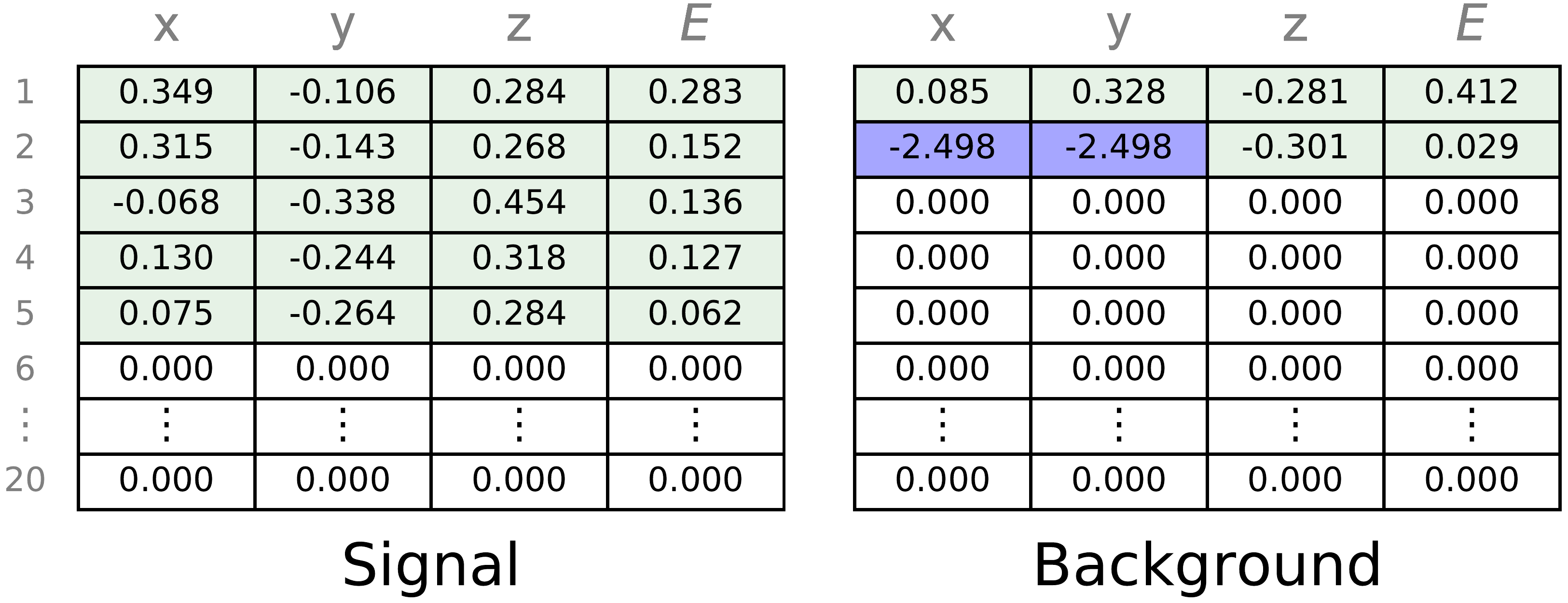}

    \caption{Example of network input for a \es excited state signal (left) and a \twonubb background event (right). The input image is fixed format of twenty rows by four columns. The four columns represent ($x, y, z, E$)  accordingly while each non-zero row represents a charge cluster. Incomplete $xy$ information are set to placeholder values (-2.498 after normalization). The energy and spatial correlations among clusters is retained by this input with all available energy and vertex information for each cluster.}
    \label{fig:input_format_of_network}
\end{figure}

The input information for the network are the energy and 3D position ($x, y, z, E$) for every reconstructed cluster of the event, as shown in Fig.~\ref{fig:input_format_of_network}. As the number of charge clusters (i.e. event multiplicity) varies event by event, the dimension of the input matrix is set to 20$\times$4, with 20 set to safely allow for the maximum number of clusters in any event. For events with multiplicity less than 20, the remaining rows are padded to zeros. The non-zero rows are arranged according to descending cluster energy order. 
The normalization of the input is done using the linear normalization formula $v_\mathrm{norm}=v/(v_\mathrm{max}-v_\mathrm{min})$, where $v$ represents the $(x, y, z, E)$ in each cluster. The maximum and minimum values are set by their corresponding physical limits in each variable so that the values of $v_{norm}$ for different variables are comparable. The normalization was found to improve training stability and lead to more separated peaks in both the signal and background discriminator distributions, though there was negligible impact on the receiver operating characteristic (ROC) curve. 

\begin{figure}[tbp!]
    \centering
    \includegraphics[width=\linewidth]{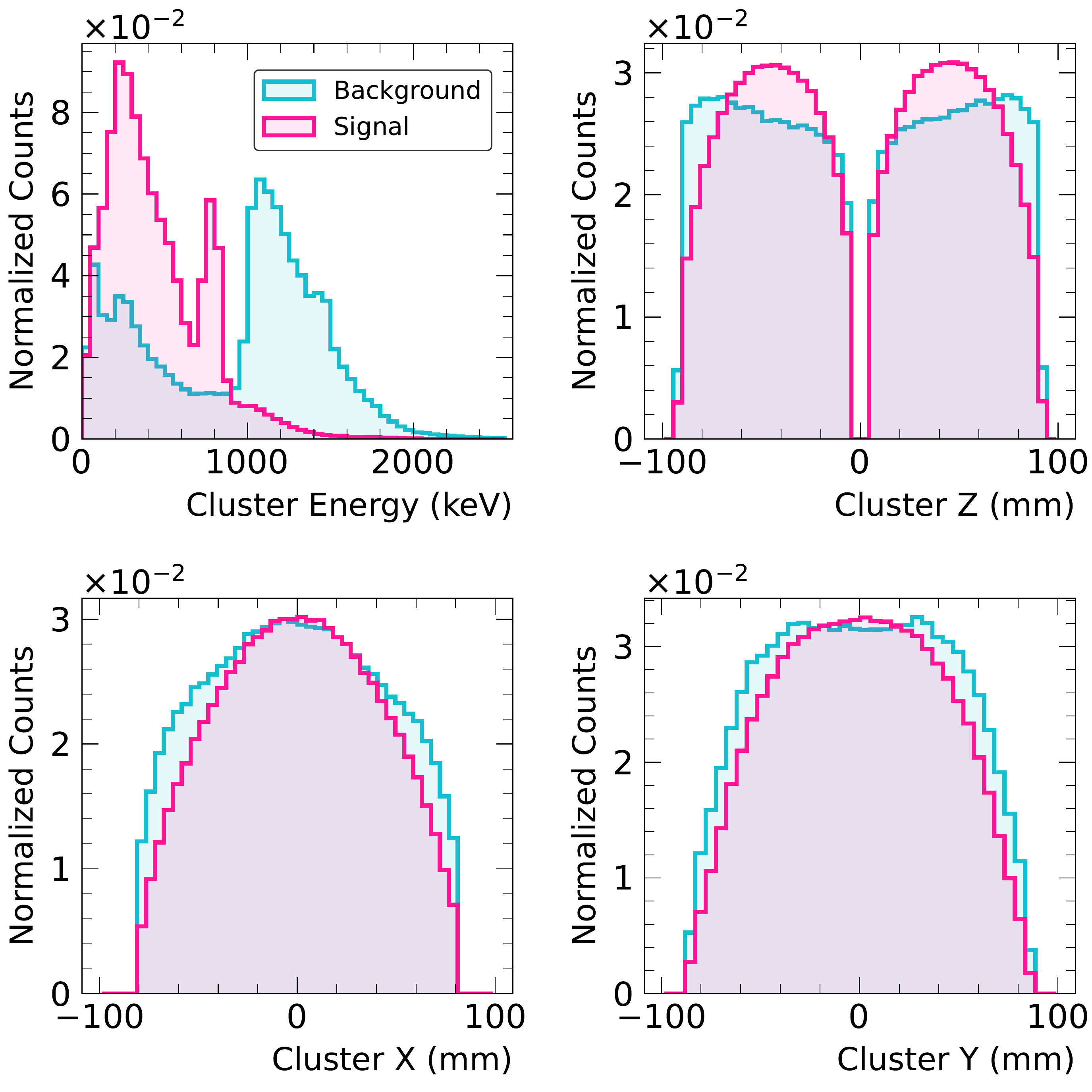}
    \caption{Distributions of the input variables to the discriminator based on MC simulations in \PII. The information of all clusters in the MS events is filled in the histogram. Useful information for background separation like number of clusters is not easily seen in 1D projections, but they are embedded in the image input.}
    \label{fig:data_distri_p2}
\end{figure}

Clusters with no reconstructed $xy$ positions are allowed in the analysis, with corresponding $xy$ values set to an unphysical placeholder values (-999) before the normalization operation mentioned above. These placeholder values were found not to affect or bias the network performance. This is validated with two networks prepared by training on two different samples: one with only full 3D events, the other also including events with incomplete $xy$ positions. The ROC area for the two networks on the same test sample of full 3D events are very similar. The ROC area of the latter network for events without full 3D positions is only slightly worse than full 3D events, presumably due to incomplete position information. The distributions of the ($x, y, z, E$) for signal and background are shown in Fig.~\ref{fig:data_distri_p2}. 

\subsection{Network structure}
\begin{figure}[tbp!]
    \centering
    \includegraphics[width=\linewidth]{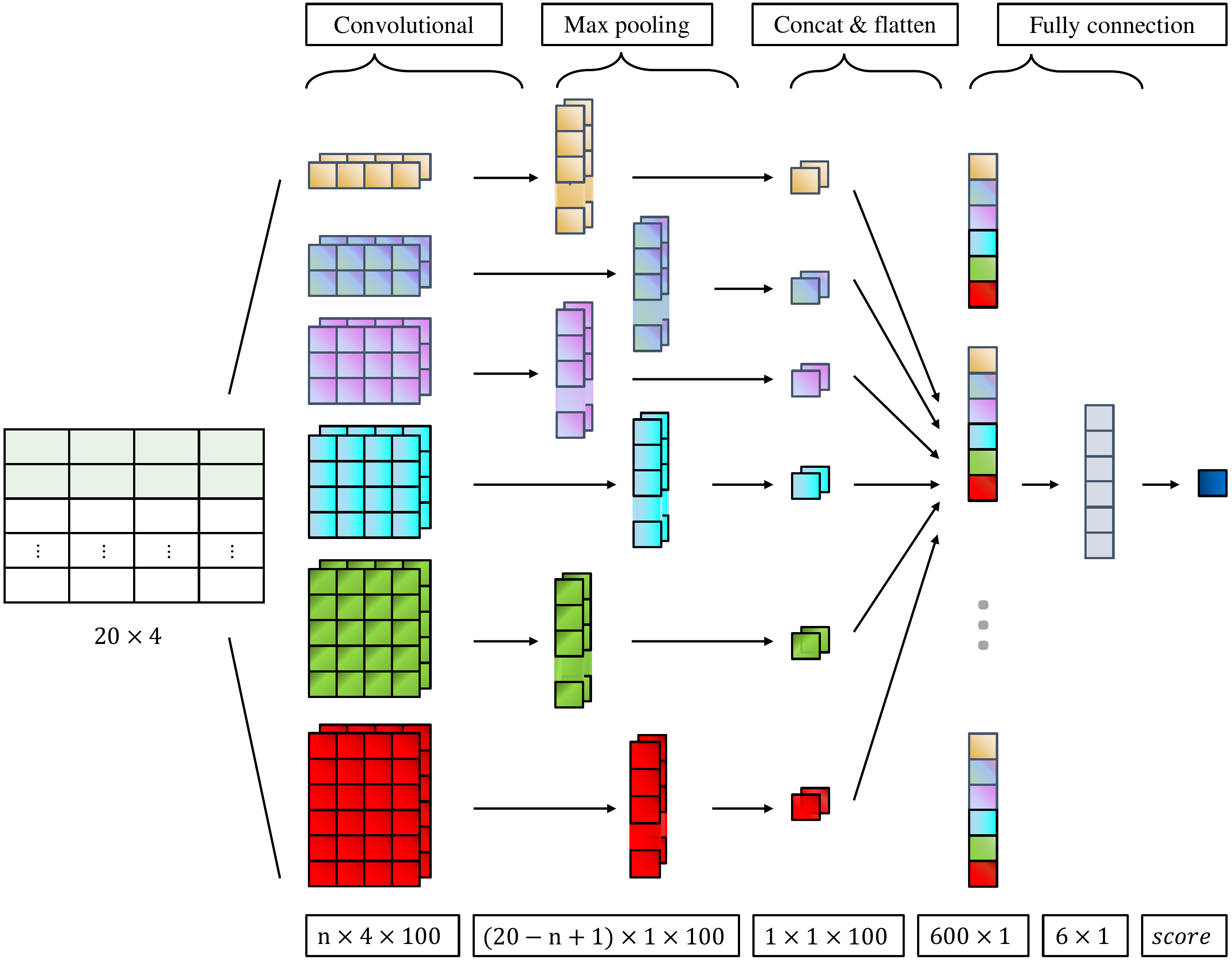}
    \caption{The CNN architecture used for background discrimination. The convolution layer consists of six different kernel sizes. The kernel shape is n$\times$4 with n ranges from 1 to 6. There are 100 convolution kernels for each size. A max pooling layer and two fully connected layers are followed.}
    \label{fig:network_structure}
\end{figure}

CNN's are one of the most commonly used DL methods in high energy physics in recent years, with applications in particle identification and event reconstruction dealing with complicated input information in the format of images. EXO-200 has used this in the \nonubb analysis for background discrimination~\cite{exo_prl_2019}, as well as analyses for event reconstruction~\cite{EXO:2018bpx_Igor}. In this analysis, a CNN with a simple structure inspired from TextCNN~\cite{zhang-wallace-textcnn} was developed to take the 20$\times$4 array of cluster information. The architecture of the network is shown in Fig.~\ref{fig:network_structure}. The network consists of a convolution layer, a max pooling layer and fully connected layers. The convolution part is composed of convolution kernels in six different sizes. The kernels have the same number of columns as the network input. The convolution is only done along the row dimension, each generating a one-dimensional array. A max pooling layer is applied, followed by two fully connected layers. The final output value after a sigmoid function gives the event a score indicating the type of event.

\subsection{Background separation performance}
\begin{figure}[htbp!]
    \centering
    \includegraphics[width=\columnwidth]{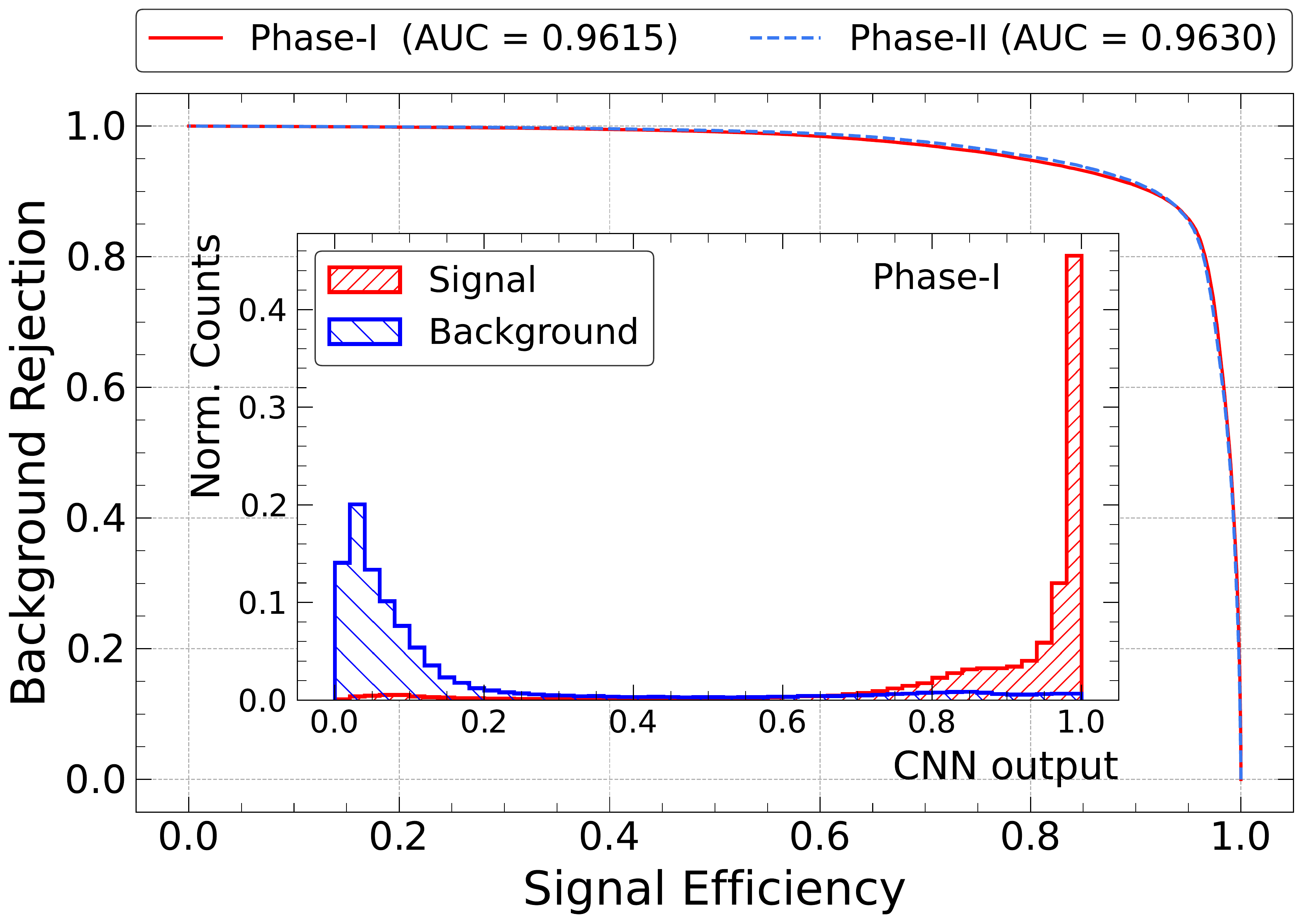}
    \caption{ROC curves of CNN discriminators in \PI and \PII. The Area Under Curve (AUC) is very similar between the two phases. The inset shows the distribution of the CNN discriminator for signal (red) and background (blue) events in \PI as an example. The background rejection efficiency shown in this plot comes from information including event topology and spectrum shape. More details on spectrum shape effect can be seen in~\cite{EXO:2018bpx_Igor,exo_prl_2019,Li:2022frp}.}
    \label{fig:comp_roc_networks_p1p2}
\end{figure}

The test sample has distributions consistent with the training sample. This implies no over-training of the network. The ROC curve, background rejection power vs signal efficiency, is shown in Fig.~\ref{fig:comp_roc_networks_p1p2}. The two phases give comparable background rejection.

\begin{figure}[htbp!]
    \centering
    \includegraphics[width=\columnwidth]{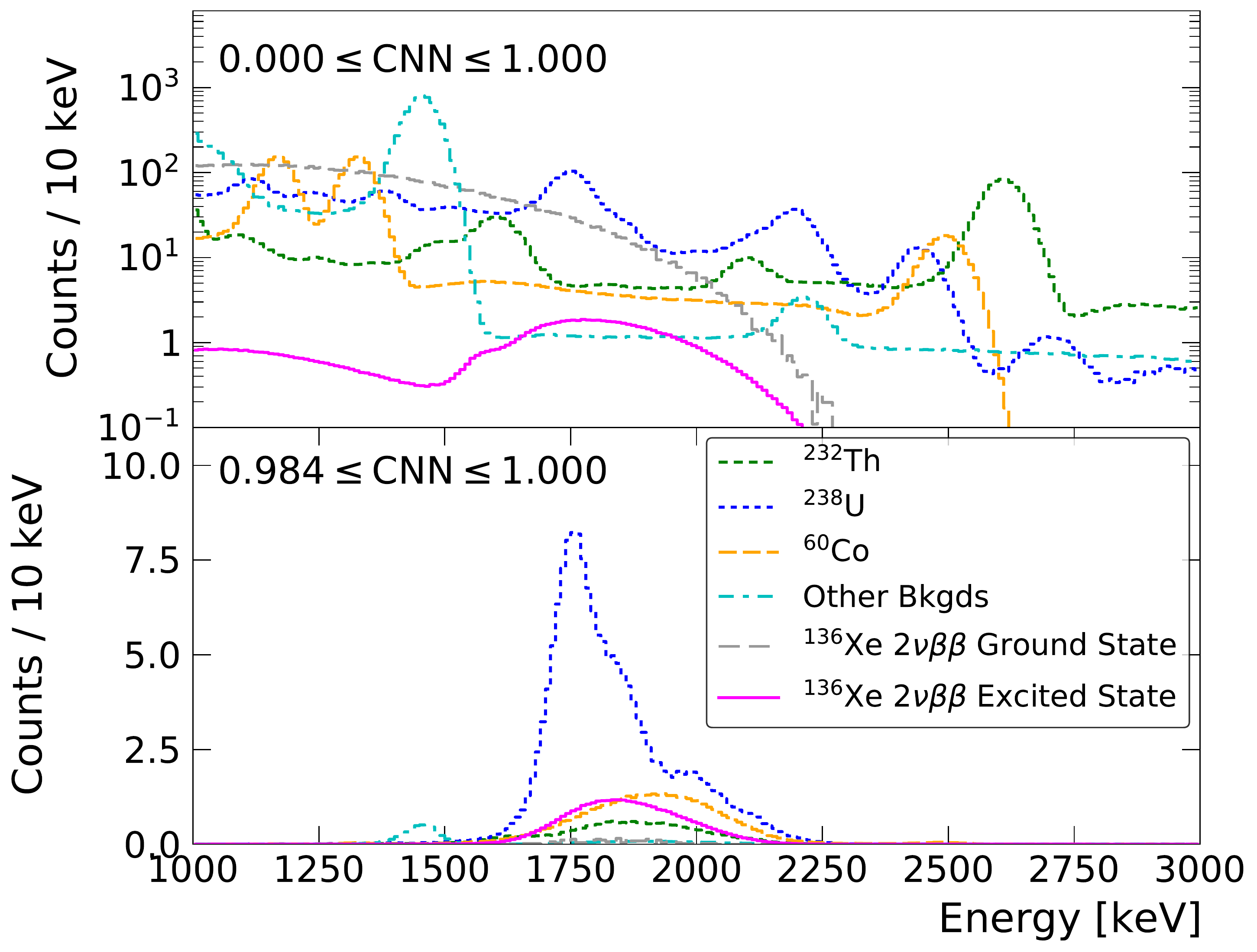}
    \caption{Background spectrum for MS events before (top) and after (bottom) a selection cut at a signal efficiency of $\sim$30\% in \PI. The simulated spectrum for signal and background components are shown. The background rates are set by the background model described in Sec.~\ref{sec:analysis overview}. For the \twonubb decay of $^{136}$Xe to excited states, the number expected from the 90\% CL sensitivity is plotted (will be discussed in Sec.~\ref{sec:results}). Other backgrounds consists of \isot{K}{40}, \isot{Co}{60}, \isot{Xe}{135}, \isot{Xe}{137} and background related to neutron capture.}
    \label{fig:residual_bkgds_cnn_p1}
\end{figure}

To understand the residual background contributions, the energy spectrum before and after a background cut on the discriminator variable is shown for illustration in Fig.~\ref{fig:residual_bkgds_cnn_p1}. \twonubb decay to the ground state of $^{136}$Ba dominates the low background data, but it can be rejected with very high efficiency using event topology information. $\gamma'$s from \isot{U}{238}, \isot{Th}{232}, \isot{K}{40} and \isot{Co}{60}, though much lower in rate, are harder to reject as they produce more clusters resembling \es state signals. An obvious energy dependence on the discrimination was observed as well. Low rejection power is found for events around 1800 keV, where the broad spectrum of the excited state peaked. As a result, \isot{U}{238}-like backgrounds remain as a dominant background after background rejection. \isot{Co}{60} is difficult to reject due to the two $\gamma$'s, leading to higher multiplicity values, like for the signals.

\subsection{Simulation and data agreement}
\label{sec:shape_agreement}
\begin{figure}[htbp!]
    \centering
    \includegraphics[width=\linewidth]{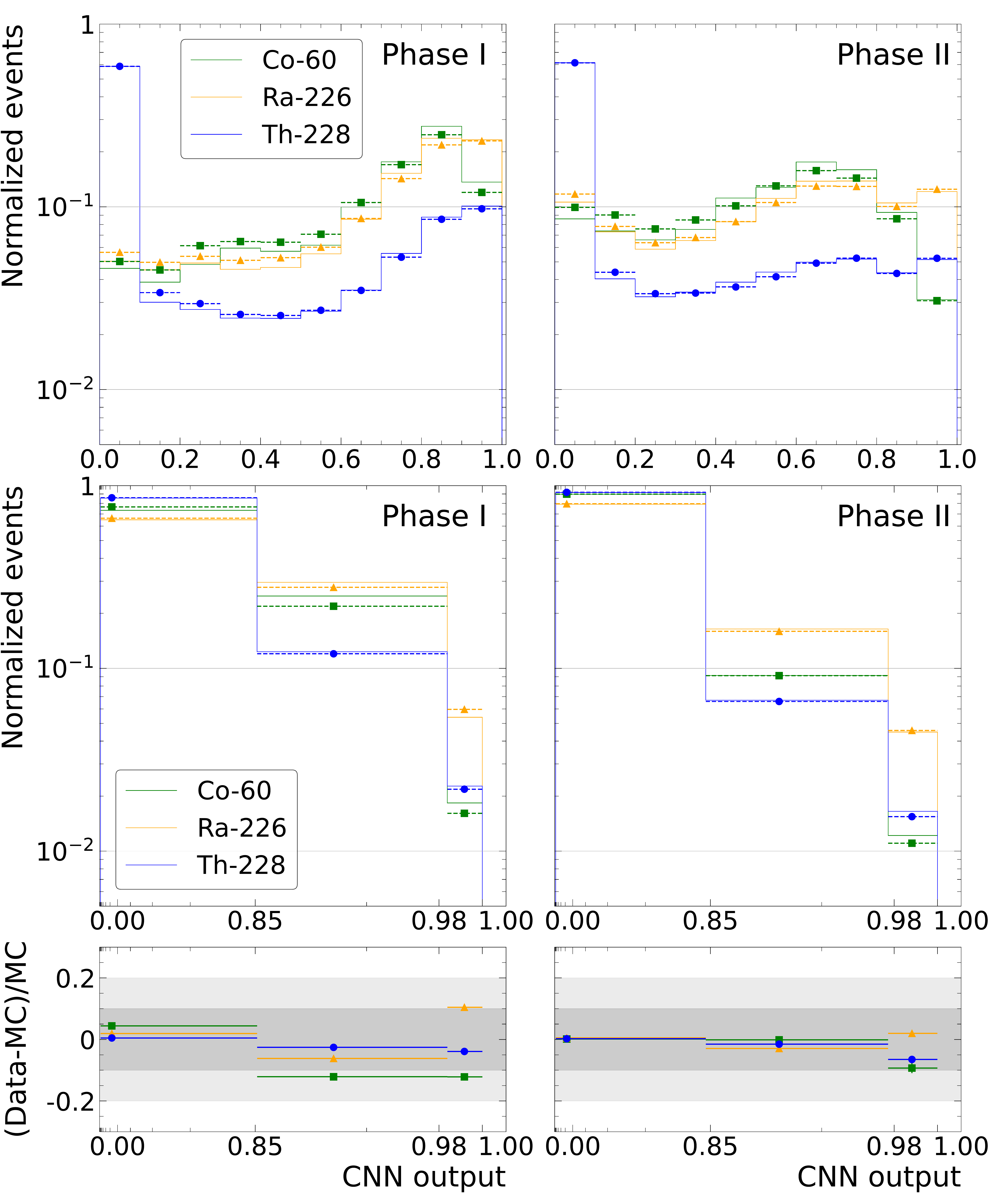}
    \caption{Shape agreement of CNN between data~(dotted line) and MC~(Solid line) for MS events using calibration sources positioned near the cathode in \PI (left) and \PII (right). The choice of binning allows equal number of excited state signal events in each bin (middle and bottom panels). The shape agreement of uniform binning is also shown in the top panel as a comparison.}
    \label{fig:shape_agree_cnn_p1p2_s5}
\end{figure}

Possible mismodeling of the spectral shape of the discriminator in Monte Carlo is a major source of systematic uncertainty. The agreement is studied by various calibration sources as shown in Fig.~\ref{fig:shape_agree_cnn_p1p2_s5}. The data is chosen to be binned with three equal efficiency bins for excited state signals. More bins are found to improve the background separation power, but with the risk of sacrificing data simulation agreement. Additionally, due to the powerful rejection ability, the background distribution decreases drastically towards a CNN value equal to 1. Therefore, fewer bins ensure enough statistics in the bin close to 1 for calibration data to constrain data simulation agreement. Data and MC agree within 15$\%$ for all source positions. The study of the binning method used only simulation and calibration data. The error introduced to the excited state signals from the residual disagreement will be discussed in Sec.\ref{sec:syst}.

\section{Systematic uncertainties}
\label{sec:syst}
The impact of systematic uncertainties is accounted for by adding Gaussian constraints to the NLL. The list of systematic uncertainties are discussed below:
\begin{enumerate}
    \item A common normalization error, caused by uncertainties in event reconstruction and selection efficiencies, is applied to all PDF components equally.
    \item SS fraction, due to uncertainties in the relative fraction of SS events out of all the events. The mean SS fractions are derived from by simulation, and their error is determined by simulation and benchmarked with calibration data. 
    \item Uncertainty in the relative fraction of neutron capture related PDF components by dedicated simulations.
    \item Uncertainty in the activity of radon in the LXe as determined in the standalone studies via measurement of $^{214}$Bi-$^{214}$Po correlated decays.
    \item Signal specific normalization error. An error associated only to signal events, allowing the signal to vary by the estimated error.
\end{enumerate}

The first four errors were evaluated in previous EXO-200 analyses~\cite{exo_prl_2019}. The common normalization errors are 3.1\% (2.9\%) for \PI (\PII), with the dominant contribution from the fiducial volume cut \cite{exo_prl_2019}. The SS fraction error is evaluated by the difference between data and MC simulation for various calibration sources at different positions. The errors are 5.8\% (4.6\%) for \PI (\PII) \cite{exo_prl_2019}. The relative capture fractions of cosmogenic neutrons is constrained with a 20\% uncertainty \cite{EXO_cosmogenics}. The radon daughters-induced background in LXe is constrained by the measured rate of \Rn decays \cite{Albert2013}.

The estimate of the signal-specific normalization error follows the same methodology as in \cite{exo_xe134}. The main difference from the result in \cite{exo_es_2016} is that a signal count dependent treatment is used in this analysis instead of a constant fractional uncertainty. This more accurately accounts for the error at small signal counts. This error varies as a function of signal counts, and consists of two main contributions: 1) shape error, caused by the level of the PDF shape agreement between data and MC; 2) the background model error due to not considering all the detailed locations of backgrounds originating from materials far from the TPC Vessel in the fit model. Instead, some representative positions are used to represent \isot{U}{238}, \isot{Th}{232}, \isot{Co}{60} from these materials. The systematic errors caused by this approximation are estimated by replacing the PDF of the remote components at different locations in the fit. The shape errors are evaluated on an ensemble of toy datasets. Each toy dataset was generated from the MC PDFs weighted by the observed data/MC ratio based on the calibration data, but fitted with the original unweighted PDFs. The difference between the injected number of signals against the fitted number of signals is taken as the shape error. The background model errors are evaluated by comparing the difference of best fit signal counts by replacing a PDF component in the background model with its alternative one at a different position. The two contributions are added in quadrature in the end, with the shape error being the dominant one. The evaluated signal normalization errors ($\sigma_{signal}$) at different injected signal numbers are found to be well described by $\sigma_{signal}/N=a/N$, with $N$ being the signal counts and $a$ being the parameter used to quantify signal-specific normalization error. The errors are summarized in Table~\ref{tab:systematics}.

\begin{table}[h]
\caption{Summary of systematic errors. The evaluated signal normalization errors at different injected signal numbers are parameterized by $\sigma_{signal}/N=a/N$, where $N$ is the signal counts.}
\label{tab:systematics}
\begin{tabular}{l m{2.0cm}<{\centering} m{2.0cm}<{\centering}}
\hline
\hline
 & \PI & \PII \\
\hline
Common normalization & 3.1\%  & 2.9\% \\
Sig-specific normalization $a$      & 30.7  & 17.9 \\
SS fraction              & 5.8$\%$  & 4.6$\%$       \\
Radon in LXe             & 10$\%$   & 10$\%$        \\
Neutron capture          & 20$\%$   & 20$\%$        \\
\hline
\hline 
\end{tabular}
\end{table}

A possible energy scale difference between beta particles and gamma particles is considered. The energy scale of beta-like events is allowed to float freely with respect to $\gamma$ events by multiplying a beta scale factor to all PDFs representing interactions of $\beta$-like events in the fit. The best fit value of beta scale is 1.0017$\pm$0.0017 (1.0008$\pm$0.0017) for \PI (\PII), suggesting a consistent energy scale within subpercent level above the 1000 keV analysis threshold.

\section{Results}
\label{sec:results}
The 90\% CL sensitivity to the half-life of the excited state decay was evaluated to be 2.0$\times10^{24}$ yr for \PI. With the improved systematic uncertainty from the CNN discriminator, optimized selection cuts and slightly larger exposure, the \PI sensitivity is improved by 15\% from the BDT-based approach in \cite{exo_es_2016}, under the new treatment of signal dependent normalization error. The new analysis of \PII data presented in this work has a slightly better sensitivity of 2.2$\times10^{24}$ yr because of smaller systematic uncertainties. Considering the current sensitivities are dominated by statistical uncertainties, the combined sensitivity can be calculated by treating the systematic errors between the two phases as independent, which gives a combined sensitivity of 2.9$\times10^{24}$ yr.

\begin{figure*}[htbp]
\centering \includegraphics[width=\linewidth]{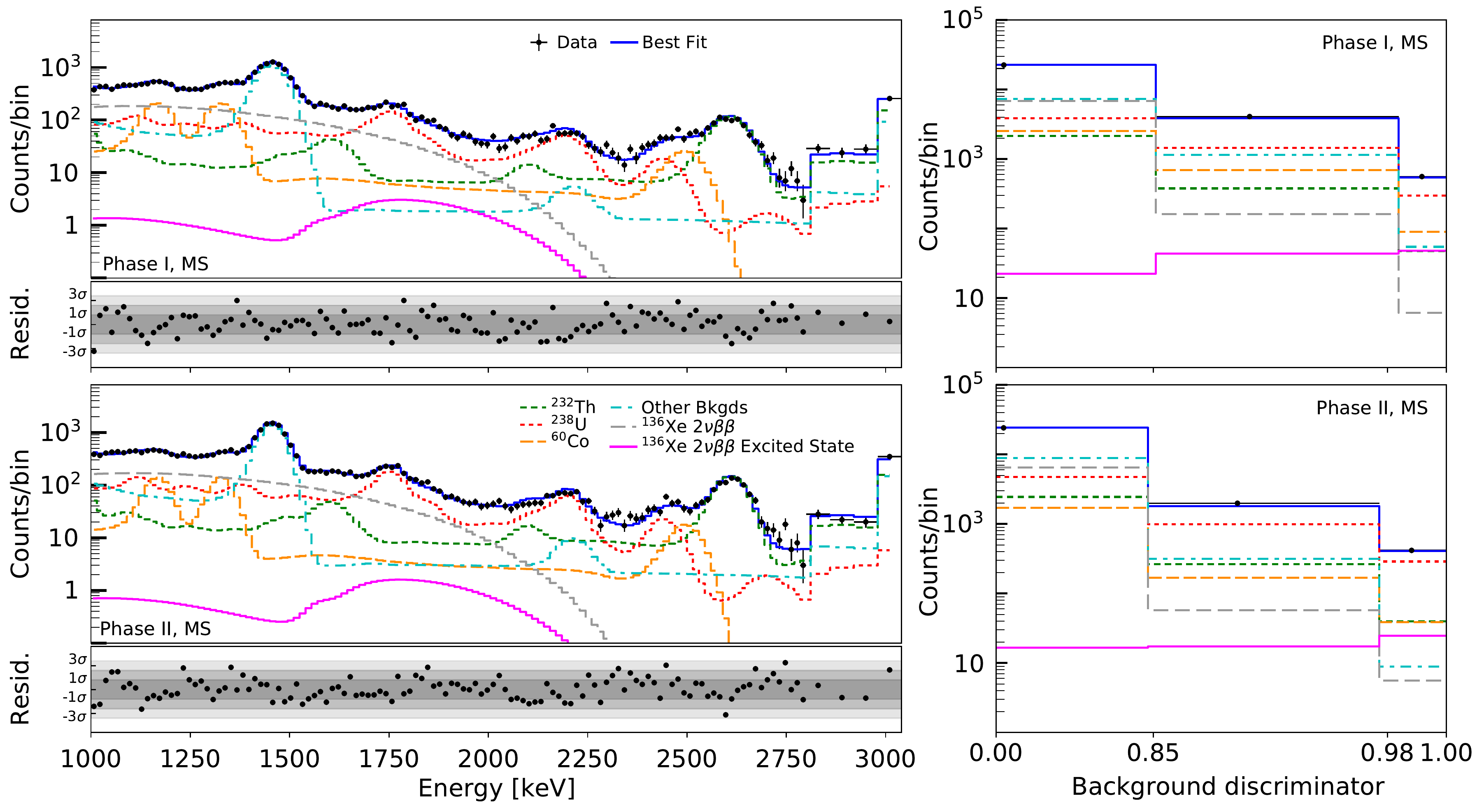}
\caption{
  Best fit to the MS low background data for energy spectrum (left) and discriminator spectrum  (right) in \PI (top) and \PII (bottom). The energy bins are 15~keV and 30~keV below and above 2800~keV, respectively.
  The best-fit residuals of the MS energy spectrum are shown for illustration, with only statistical uncertainty taken into account. The small deviations are taken into account in the spectral shape systematic errors.
}
\label{fig:p1p2_energy_ml_spectrum_fit}
\end{figure*}

A final fit using energy and CNN as fitting dimensions was applied to the full EXO-200 dataset. We found no statistically significant signals in either phase (Fig.~\ref{fig:p1p2_energy_ml_spectrum_fit}). A lower limit on the half-life is obtained to be 0.9$\times10^{24}$ yr and 1.4$\times10^{24}$ yr for \PI and \PII.
The combination of the two phases gives a limit of 1.4$\times10^{24}$ yr.
While there are large uncertainties from different nuclear models, the result in this work is in tension with the values predicted by QRPA as summarized in Table~\ref{tab:t12_136Xeexc}. The future nEXO experiment~\cite{nEXOpCDR} is expected to greatly improve the search capabilities since it is expected to fully contain the de-excitation gammas and allow lower backgrounds, as well as have much more exposure~\cite{nEXO:2021_nEXO_sens}. 

\begin{table}[t]
\caption{Theoretical and experimental results of $^{136}{\rm Xe}$ $2\nu\beta\beta$-decay half-life to the $0^+_1$ state of $^{136}{\rm Ba}$.}
\label{tab:t12_136Xeexc}
    \centering
    \begin{tabular}{ll m{1.5cm}<{\centering} m{2.5cm}<{\centering}}
        \hline
        \hline
          &  & Reference & $T^{2\nu}_{1/2}$ ($10^{23}$~yr) \\ \hline
         
        \multirow{6}*{Theory} & QRPA  & \cite{Jokiniemi:2022_es_theory_latest} & $0.14-13$  \\
         & QRPA & \cite{theory_predict} & $1.3-8.9$ \\ 

         & IBM-2 & \cite{Jokiniemi:2022_es_theory_latest} & $(1.5-3.6)\cdot 10^{2}$\\
         & IBM-2 & \cite{Barea:2015kwa} & $2.5 \cdot10^2$\\  
         & NSM   & \cite{Jokiniemi:2022_es_theory_latest} & $(2.5-6.6)\cdot10^3$\\  
         & EFT& \cite{Jokiniemi:2022_es_theory_latest} & $(0.62-16)\cdot 10^{2}$\\ \hline 
         \multirow{3}*{Experiment} & KamLAND-Zen   & \cite{KamLAND-Zen:2015tnh_kamland_es} & $>8.3$\\ 
         & EXO-200 (2016)  & \cite{exo_es_2016} & $>6.9$ \\
         & EXO-200   & This work & $>14$\\ 
        \hline
        \hline
    \end{tabular}
\end{table}

\section{Conclusion}
The results of a refined search for \twonubb of $^{136}$Xe to the \es excited state of $^{136}$Ba using EXO-200 are reported in this paper. No statistically significant evidence for this process is found and a limit on the half-life of $T_{1/2}>1.4\times10^{24}$~yr at 90\% CL is obtained. 
A CNN based discriminator was utilized in this analysis, which fully exploits the cluster information  for background rejection while achieving good agreement between data and simulation, leading to an improvement by a factor of 1.7 relative to the current best constraint set by KamLAND-Zen. Future ton-scale experiments \cite{nEXOpCDR,KamLAND-Zen:2022tow} might have a chance to make a discovery. 

\begin{acknowledgments}
EXO-200 is supported by DOE and NSF in the United States, NSERC in
Canada, SNF in Switzerland, IBS in Korea, DFG in
Germany, and CAS in China. EXO-200 data analysis and
simulation uses resources of the National Energy Research Scientific
Computing Center (NERSC).  We gratefully acknowledge the KARMEN
collaboration for supplying the cosmic-ray veto detectors, and the
WIPP for their hospitality.
\end{acknowledgments}

\onecolumngrid

\twocolumngrid

\bibliography{exo_es_search}
\end{document}